# A Compact Ultra-Wideband Circularly Polarized Antenna Based on Miniaturized Phase Shifter

Han-Jie Xu, Shi-Wei Qu, *Senior Member, IEEE*

*Abstract*—In this article, a compact wideband circularly polarized antenna based on a miniaturized phase shifter with ultra-wideband operation is proposed. The proposed antenna is comprised of a pair of compact orthogonal ultra-wideband Vivaldi antennas and a miniaturized phase shifter. To achieve wideband impedance matching and miniaturization, parasitic radiation structures, metal coupled plates, and Γ-type balun with high-impedance transmission lines are designed. After optimization, the final antenna dimensions are only $0.36 \times 0.36 \times 0.34\lambda_l^3$, where $\lambda_l$ is the free-space wavelength at the lowest operating frequency. Additionally, the miniaturized 90° wideband phase shifter of the antenna is designed by employing a Π-type network and a negative group delay (NGD) network with extremely compact dimensions of $0.071 \times 0.047\lambda_l^2$. The simulated results indicate that the antenna exhibits a 10-dB impedance bandwidth within 0.32 ~ 1.2 GHz (3.75:1) and a 3-dB axial ratio (AR) bandwidth within 0.32 ~ 1.15 GHz (3.59:1). Finally, a prototype is fabricated and reasonable agreement is achieved between the simulated and measured results.

## I. INTRODUCTION

Circularly polarized (CP) antennas have been widely studied and applied in the last decades due to their advantage of multipath immunity. Recently, with the increasing demands on modern wireless systems of large channel capacity and miniaturization, the designs of the wideband, compact, and low-profile CP antennas are of great significance and attract the attention of researchers.

CP antennas with single-feed designs are one of the promising solutions to meet the aforementioned targets, owing to their low profiles and low manufacturing cost [1]-[4]. For instance, in [3], a particular feed point is selected to attain circular polarization with a low axial ratio (AR). The antenna achieves a profile of $0.11\lambda_l$, featuring a simple structure, but exhibits a relatively narrow 3-dB AR bandwidth of only 2.32 ~ 2.95 GHz (23.9%). Additionally, various types of single-feed CP antennas with simple structures have been reported, including magnetoelectric dipole antennas [5]-[8], dielectric resonator antennas [9]-[10], monopole antennas [11]-[12], and wide-slot antennas [13]-[14]. However, these antennas often exhibit limited operating bandwidths, which are insufficient to meet the requirements of modern wireless systems.

Compared to single-feed CP antennas, integrating a multi-feed network within an ultra-wideband antenna is an effective approach to broaden the AR bandwidth. Moreover, Vivaldi antennas have been extensively utilized in the design of ultra-wideband CP antennas due to their simple structure, large bandwidth, and favorable radiation characteristics. For example, four sequentially rotated Vivaldi elements loaded with dielectric lenses achieve a 3-dB AR bandwidth from 11.22 to 40.53 GHz (3.61:1), in [15], but result in a high profile ($4.65\lambda_l$). To reduce the profile, a halved-type Vivaldi antenna is reported in [16], but its additional circular metal wall limits its application. Considering this issue, a novel halved-type Vivaldi antenna is designed in [17], which serves as both the radiator and the ground plane. Although it has a 3-dB AR bandwidth within 0.49 ~ 1.47 GHz (3:1), its profile ($0.61\lambda_l$) is higher compared to the antenna in [16]. Similarly, a 90° phase shifter and a pair of orthogonal Vivaldi antennas are integrated on the vertical printed circuit boards (PCBs), as reported in [18], achieving a 3-dB AR bandwidth of 2.05 ~ 8 GHz (3.9:1). However, the large sizes of the phase shifter result in the antenna having a high profile ($0.63\lambda_l$). While these CP antennas with Vivaldi elements exhibit a large AR bandwidth, high profile limits their practical applications.

To address the previously mentioned issue, various low-profile designs for CP antennas have been proposed. In [19], the novel CP antenna with bent Vivaldi antenna elements has a profile of only $0.12\lambda_l$, but with a relatively narrow AR bandwidth within 1.65 ~ 2.77 GHz (1.68:1). Furthermore, CP metasurface antennas also exhibit the advantages of wideband and low profile. In [20], a CP metasurface antenna array with sequential rotation feeding shows a bandwidth from 3.3 to 7.75 GHz (2.35:1) and a profile of $0.05\lambda_l$, but the dimensions ($0.8 \times 0.8\lambda_l^2$) are relatively large. In addition, a modified Schiffman phase shifter and a meandering L-type microstrip feed are employed in a CP patch antenna described in [21]. Although the profile is only $0.05\lambda_l$, it also has the dimensions as large as $0.87 \times 0.87\lambda_l^2$. In order to minimize the antenna dimensions, a multimode operation CP patch antenna is proposed in [22]. The AR bandwidth of the antenna is 2.2 ~ 5.5 GHz (2.5:1), and the patch dimensions are only $0.29 \times 0.29\lambda_l^2$. However, the dimensions of the complicated phase shifter are nearly twice that of the patch. In summary, the aforementioned antennas have either a narrow bandwidth, a high profile, large antenna

This work was partly supported by the National Natural Science Foundation of China (NSFC) Projects under No. U20A2016. (Corresponding to *Shi-Wei Qu*).
The authors are with the School of Electrical Science and Engineering, University of Electronic Science and Technology of China (UESTC), Chengdu 611731, China (e-mail: shiweiqu@uestc.edu.cn; hanjiexu@std.uestc.edu.cn).

dimensions, or large phase shifter dimensions.

Considering these issues, a novel compact ultra-wideband CP antenna based on a miniaturized phase shifter is proposed in this paper. The metal coupled plates and sawtooth structures are employed to enlarge the surface current path, thereby moving the lower operating frequency downwards. Meanwhile, a parasitic radiation structure and Γ-type balun with high-impedance transmission lines are utilized to achieve broadband impedance matching. Ultimately, the 10-dB impedance bandwidth of the proposed antenna is 0.32 ~ 1.2 GHz (3.75:1), and the 3-dB AR bandwidth is 0.32 ~ 1.15 GHz (3.59:1), with dimensions of only $0.36 \times 0.36 \times 0.34\lambda_l^3$. Moreover, a novel miniaturized ultra-wideband phase shifter based on a Π-type network and a negative group delay (NGD) network is proposed. Its bandwidth is 0.3 ~ 1.1 GHz (3.67:1), and the dimensions are only $0.071 \times 0.047\lambda_i^2$, representing a substantial reduction in dimensions compared to conventional phase shifters utilized in CP antennas.

## II. ANTENNA OVERVIEW

The overall structure proposed in this paper can be divided into two parts: compact Vivaldi antenna and feed network, as shown in Fig. 1. The Vivaldi antenna is printed on an F4BM-300 substrate with a relative dielectric constant of 3 and a thickness of 0.8mm, as illustrated in Fig. 1(a). In addition to conventional tapered slots, a parasitic radiation structure, sawtooth structures and an Γ-type balun with high-impedance transmission lines are employed, as shown in Fig. 1(b). Meanwhile, four metal coupled plates are arranged around the antenna to facilitate impedance matching at low frequencies.

Apart from the vertical Vivaldi antenna, the feeding network is printed on an F4BM-220 substrate with a relative dielectric constant of 2.2 and a thickness of 1mm. It consists of three components: a Π-type network, an NGD network, and an unequal Wilkinson power divider, as shown in Fig. 1(c). On one hand, the Π-type network is composed of two parallel inductors and one series capacitor, as shown in Part A of Fig. 1(c). On the other hand, the NGD network consists of two parallel RLC resonators, a series inductor, and two tapered transmission lines, as shown in Part B of Fig. 1(c). However, the utilization of resistive elements leads to increased loss within the NGD network. To achieve equal power division, an unequal Wilkinson power divider is adopted, as shown in Part C of Fig. 1(c).

The overall dimensions of the proposed antenna are $0.36 \times 0.36 \times 0.34\lambda_l^3$, and the detailed geometric parameters are shown in Fig. 1. Additionally, the dimensions of the proposed phase shifter are $0.071 \times 0.047\lambda_i^2$, representing a significant reduction compared to the conventional 90° phase shifter utilized in CP antennas. After optimization, the proposed antenna exhibits favorable performance with the simulated results shown in Fig. 2. The 10-dB impedance bandwidth of the proposed antenna is 0.32 ~ 1.2 GHz (3.75:1), and the 3-dB AR bandwidth is 0.32 ~ 1.15 GHz (3.59:1) with a peak gain of 6.6 dBic at 0.85 GHz. Furthermore, the design of the antenna and feed network will be discussed in detail in Sections III and IV, respectively.

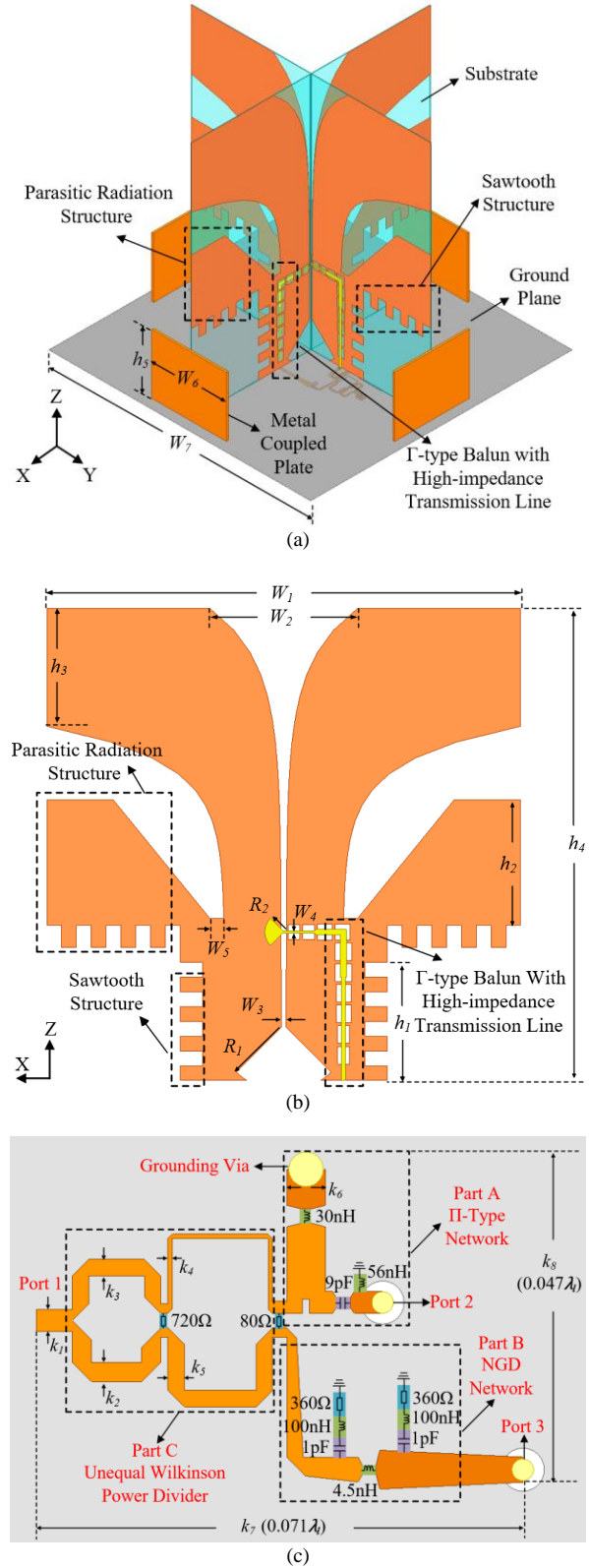

Fig. 1. Proposed antenna topology with key parameters. (a) Trimetric view. (b) Side view of the Vivaldi structure. (c) 90° phase shifter. Key parameters: $W_1$=320, $W_2$=100, $W_3$=3.4, $W_4$=0.8, $W_5$=9.8, $W_6$=100, $W_7$=350, $h_1$=84.6, $h_2$=85, $h_3$=80, $h_4$=320, $h_5$=80, $k_1$=3, $k_2$=2.7, $k_3$=2.3, $k_4$=0.6, $k_5$=2.3, $k_6$=5.4, $k_7$=67, $k_8$=44.5 (in mm).

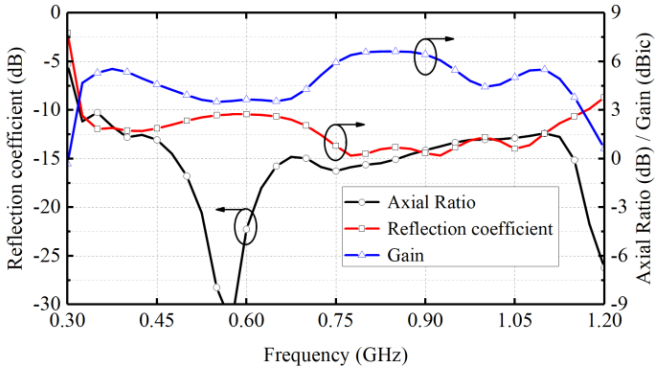

Fig. 2. Simulated reflection coefficient, AR, and Gain of the proposed antenna.

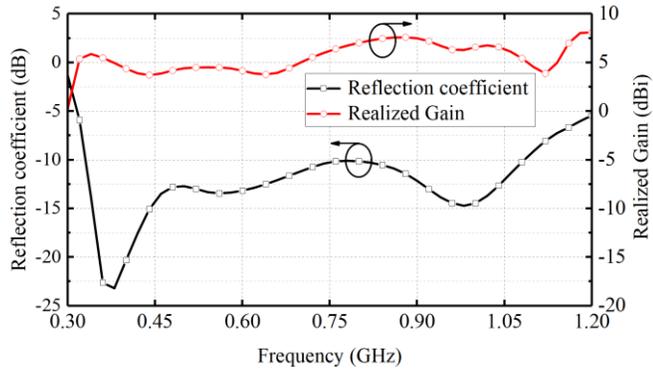

Fig. 3. Simulated reflection coefficient and Realized Gain of the Vivaldi antenna.

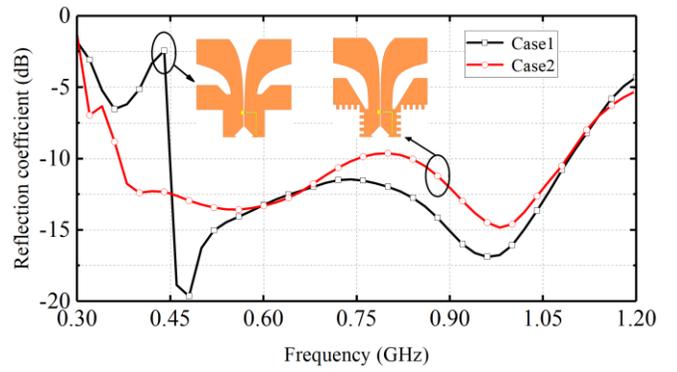

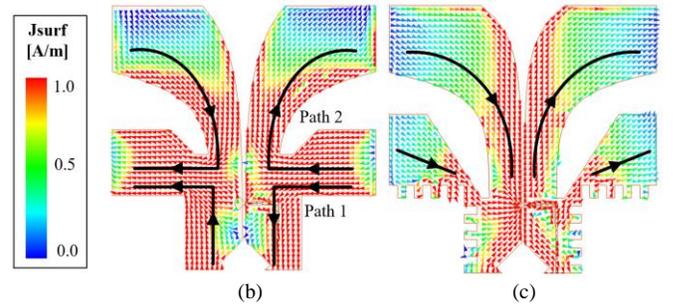

Fig. 4. Simulated results of two different cases. (a) reflection coefficients of two different cases. (b) surface current distribution of Case1 at 0.44 GHz and (c) surface current distribution of Case2 at 0.44 GHz.

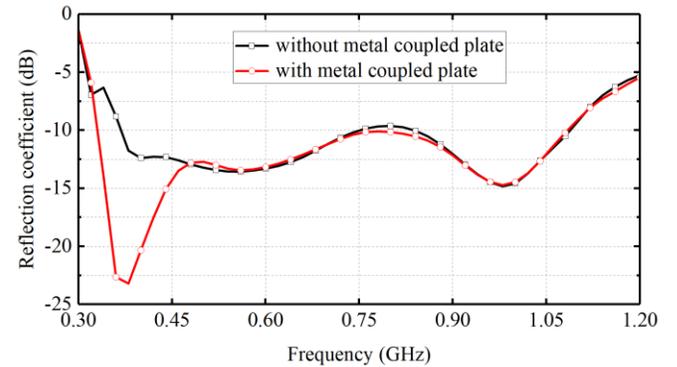

Fig. 5. Simulated reflection coefficients of the proposed Vivaldi antenna with and without metal coupled plate.

## III. ANTENNA DESIGN

Different from conventional Vivaldi antennas, the proposed antenna utilizes a parasitic radiation structure instead of conventional high tapered slots. This approach not only moves the lower operating frequency downwards but also facilitates the miniaturization. However, the use of a parasitic radiation structure may lead to undesired resonance occurring within the operating bandwidth. To address this issue, sawtooth structures are employed to shift the resonance to lower frequencies. Additionally, metal coupling plates and the Γ-type balun with high-impedance transmission lines are crucial for achieving wideband and miniaturization. As illustrated in Fig. 3, the proposed Vivaldi antenna achieves an impedance bandwidth from 0.32 GHz to 1.08 GHz, with overall sizes of $0.36 \times 0.36 \times 0.34\lambda_l^3$.

To further clarify the undesired resonances within the operating frequency band, Fig. 4(a) shows the reflection coefficients before and after introducing the sawtooth structures and modified parasitic radiation structure (Case1 and Case2). The results reveal a distinct resonance at 0.44 GHz in Case1. At this moment, the current distributions of the tapered slot are opposite to that of the parasitic radiation structure as shown in Fig. 4(b), resulting in the cancellation of far-field radiation. The reason for the reverse current is the existence of two distinct resonant current paths, Path1 and Path2, on the surface of the Vivaldi element. In order to shift the resonance towards a lower frequency, modified parasitic radiation structure and sawtooth structures are employed in Case2, as illustrated in Fig. 4(a), respectively enlarging the current paths of Path2 and Path1. Subsequently, the resonance is shifted beyond the operational band and the direction of currents on the tapered slot and the parasitic radiation structure is consistent, as shown in Fig. 4(a) and Fig. 4(c). Consequently, sawtooth structures and the modified parasitic radiation structure can not only broaden the bandwidth but also facilitate the miniaturization of the antenna.

Additionally, the metal coupled plates positioned around the antenna also facilitate impedance matching at low frequencies. Through the capacitance formed between the metal coupled plates and the parasitic radiation structure, the current path of Path1 and Path2 is further enlarged, thereby shifting the resonance to lower frequencies. As shown in Fig. 5, the reflection coefficient of the antenna with metal coupled plates loaded achieves a satisfactory improvement in the low frequencies. Moreover, the design of Γ-type balun with high-impedance transmission lines is important for attaining a large bandwidth. Compared to conventional Γ-type baluns, the balun with a high-impedance transmission line provides greater design flexibility. As shown in Fig. 6, the reflection coefficient after loading has been improved within the operating bandwidth.

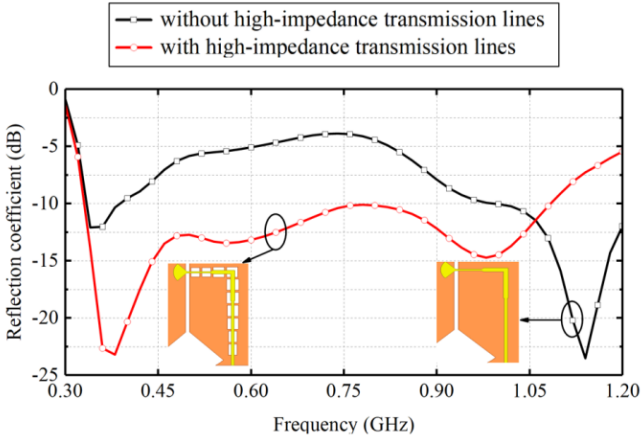

Fig. 6. Simulated reflection coefficients of the proposed Vivaldi antenna with and without high-impedance transmission lines.

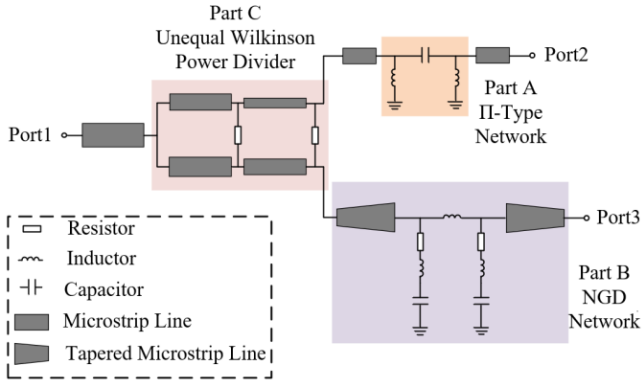

Fig. 7. Schematic of the ultra-wideband miniaturized 90° phase shifter.

## IV. FEED NETWORK DESIGN

A broadband feeding network is necessary for a dual-feed CP antenna. As shown in Fig. 7, a novel wideband miniaturized 90° phase shifter is designed, with specific parameters detailed in Fig. 1(c). The feed network can be segmented into three parts: Part A, Part B, and Part C. Part A, serving as the main line, incorporates a Π-type network consisting of two parallel inductors and one series capacitor. Part B, acting as the reference line, includes two tapered microstrip lines and an NGD network, which comprises two parallel RLC resonators and a series inductor. To account for the insertion loss of the NGD network, Part C is composed of a two-stage unequal Wilkinson power divider. Fig. 8 shows that the dual-feed network has an extremely large 10-dB impedance bandwidth from 0.3 to 1.1 GHz (3.67:1), where the insert loss is below 1.1 dB, the amplitude difference is no more than 0.6 dB and the phase deviation is less than 10°. This innovative design achieves a substantial reduction in sizes compared to traditional phase shifters utilized in CP antennas.

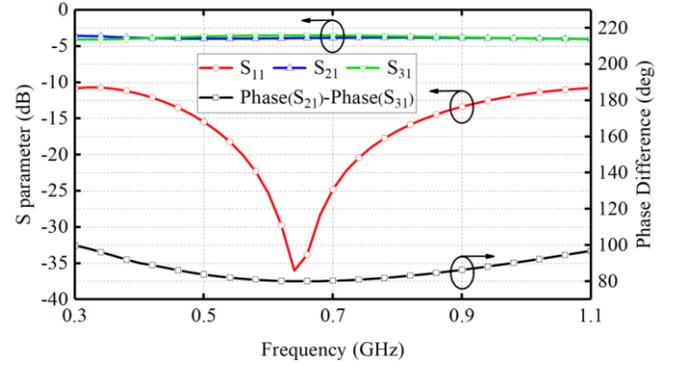

Fig. 8. Simulated results of the designed 90° phase shifter.

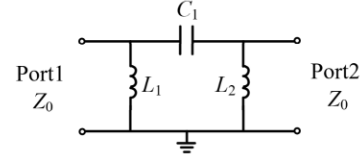

Fig. 9. Core equivalent circuit model of the Π-network.

### A. Design of the Π-type network

Traditional 90° phase shifters commonly utilize a coupled line as the main line and a long 50 Ω transmission line as the reference line. Due to the principle of the coupled line, it exhibits a large lagging phase($S_{21}$) at low frequencies, which necessitates a long reference line to achieve a 90° phase difference. To address this concern, a Π-type network is designed to replace the coupled line, which provides a leading phase($S_{21}$) at low frequencies.

As shown in Fig. (9), the core equivalent circuit of the Π-type network is a two-port network consisting of two inductors connected to ground and in series with a capacitor, with specific structure and parameters depicted in Fig. 1(c). Therefore, its corresponding normalized ABCD matrix is as follows:

$$[A] = \begin{bmatrix} 1 & 0 \\ -jY_1 & 1 \end{bmatrix} \begin{bmatrix} 1 & -jZ_1 \\ 0 & 1 \end{bmatrix} \begin{bmatrix} 1 & 0 \\ -jY_2 & 1 \end{bmatrix}$$
$$= \begin{bmatrix} 1 - Y_2 Z_1 & -jZ_1 \\ -jY_1 - jY_2 + jY_1 Y_2 Z_1 & 1 - Y_1 Z_1 \end{bmatrix} \quad (1)$$

where the relationship among $C_1$, $L_1$, $L_2$, $Z_1$, $Y_1$, $Y_2$, Port impedance $Z_0$ and angular frequency $\omega$ is given as

$$Y_1 = \frac{Z_0}{\omega L_1} \quad Y_2 = \frac{Z_0}{\omega L_2} \quad \text{and} \quad Z_1 = \frac{1}{Z_0 \omega C_1} \quad (2)$$

The transmission coefficient of the Π-type network can be expressed by the ABCD matrix

$$S_{21} = \frac{2}{A + B + C + D}$$
$$= \frac{2}{2 - (Y_1 + Y_2)Z_1 - j(Y_1 + Y_2 + Z_1 - Y_1 Y_2 Z_1)} \quad (3)$$

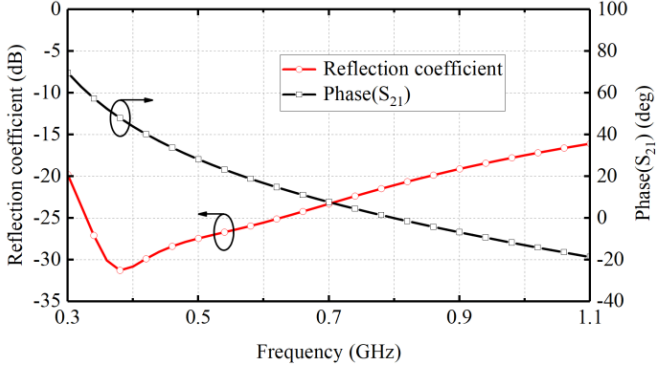

Fig. 10. Simulated reflection coefficient and phase($S_{21}$) of the Π-type network.

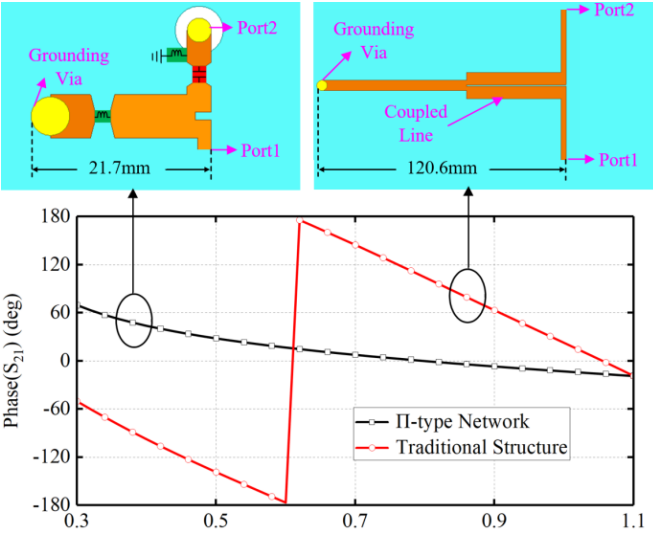

Fig. 11. Simulated phase($S_{21}$) of the traditional structure and the Π-type network.

Additionally, the phase of $S_{21}$ and its corresponding group delay can be naturally calculated as

$$\angle S_{21} = \arctan(\frac{Y_1 + Y_2 + Z_1 - Y_1 Y_2 Z_1}{2 - (Y_1 + Y_2) Z_1}) \quad (4)$$

$$\tau_{delay} = -\frac{\partial \angle S_{21}}{\partial \omega} \quad (5)$$

As mentioned above, the capacitance and inductance values can be adjusted to tune the phase($S_{21}$) and the impedance of the Π-type network. As shown in Fig. 10, the reflection coefficient of the Π-type network is below -15 dB, exhibiting a leading phase characteristic at low frequencies with a loss of less than 0.1 dB. To further demonstrate the advantages of the Π-type network, a comparison of the phase($S_{21}$) between the traditional coupled line and the Π-type network is presented in Fig. 11. For instance, at 0.3 GHz, the phase($S_{21}$) of the coupled line structure is -50°, while that of the Π-type network is 70°. To achieve a 90° phase difference, the phase of the reference lines should be -140° and -20°, respectively. To rephrase, the coupled line necessitates a reference line length of 0.39$\lambda_l$, whereas the Π-type network requires only 0.056$\lambda_l$. Moreover, in comparison with the traditional coupled line, the Π-type network itself achieves an 82% reduction in sizes.

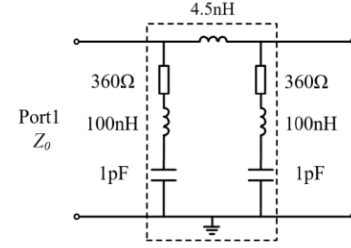

Fig. 12. Core equivalent circuit model of the NGD network.

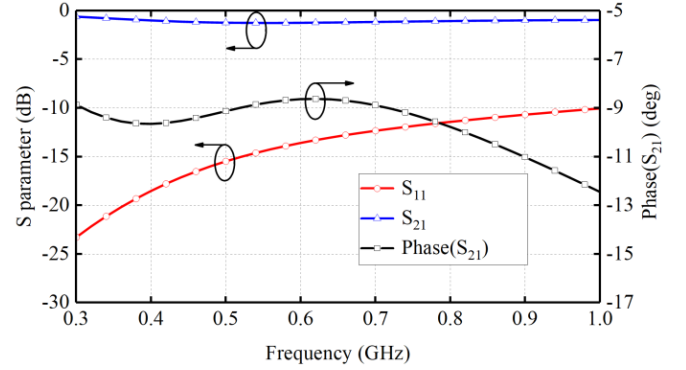

Fig. 13. Simulated S-parameters and phase($S_{21}$) of the NGD network.

### B. Design of the NGD network

In addition to utilizing the Π-type network as the main line, adopting the NGD network as a reference line is essential for achieving wideband and miniaturization. In traditional designs, realizing a 90° phase difference often requires a long reference line, leading to a significant group delay and difficulty in matching the group delay of the main line. This issue hinders the stable phase difference within a large bandwidth. Recognizing this concern, the NGD circuit is employed to reduce both the group delay and the length of the reference line. By serially and in parallel connecting RLC resonators, this circuit achieves negative group delay. While a complicated circuit is employed in [25] to achieve excellent performance, the high losses necessitate the use of an amplifier for compensation, resulting in the circuit to unidirectional transmission. To balance the group delay and losses, a simplified NGD circuit is designed as shown in Fig 12. This circuit provides a phase($S_{21}$) variation between -8.8° and -12.6° within the frequency range of 0.3 GHz to 1 GHz. Meanwhile, the reflection coefficient is below -10 dB, as depicted in Fig. 13.

To further elucidate the advantages of the circuit, it is integrated into the transmission line, as illustrated in Fig 14. When only loading an 83 mm length of the transmission line, the phase($S_{21}$) is -30° at 0.3 GHz and -100° at 1 GHz, as shown in Fig. 15. Integrating the NGD circuit into the 83 mm length of the transmission line makes an overall phase lag of 10° compared to the unloaded NGD circuit, while the group delay remains essentially unaltered. Meanwhile, extending the transmission line to 111 mm also achieves a 10° phase lag at 0.3 GHz compared to 83mm, while a 37° phase lag is observed at 1 GHz. Thus, extending the transmission line results in significant group delay, making it difficult to achieve stable phase difference within a large bandwidth. In addition, as shown in Fig. 13, the NGD circuit exhibits a loss of less than 1.27 dB, but it is only implemented on the single path of the phase shifter. Considering the losses introduced by the NGD circuit, an unequal Wilkinson power divider is employed for balancing, as shown in Fig. 1(c). Moreover, owing to the compact dimensions of $0.071 \times 0.047\lambda_l^2$, the final phase shifter exhibits minor transmission loss. Therefore, the overall loss of the Π-type network, NGD network, and unequal Wilkinson power divider cascade is less than 1.1 dB, as shown in Fig. 8. In summary, the NGD circuit not only achieves broadband stable phase differences by decreasing the group delay but also reduces the sizes of the phase shifter.

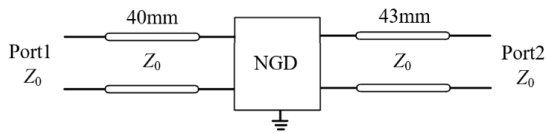

Fig. 14. Schematic of NGD circuit loaded with transmission lines.

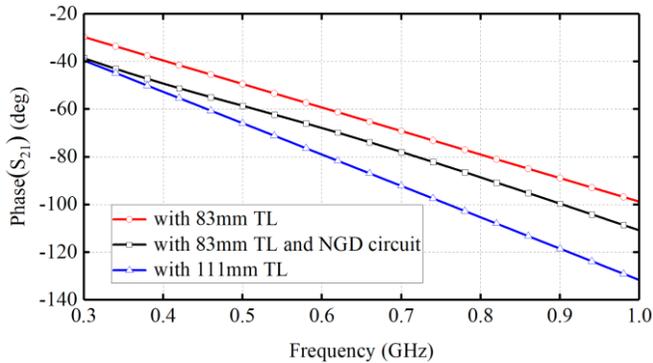

Fig. 15. Simulated phase($S_{21}$) of 83 mm transmission line, 83 mm transmission line with NGD circuit and 111 mm transmission line.

## V. Measurement and Validation

To validate the performance of the proposed antenna, a prototype was fabricated and measured, as depicted in Fig. 16. The prototype comprises two main components: the compact ultra-wideband Vivaldi elements and a miniaturized feed network. A pair of orthogonal Vivaldi elements are mounted on the ground plane with the aid of stands, and their effective grounding is ensured by copper foil. Metal coupled plates around the proposed antenna are mounted to the ground plane and also serve to stabilize the antenna. The feeding network is located below the ground plane, with lumped elements soldered onto it correspondingly. An ABS substrate with a relative dielectric constant of 3 is placed below the ground plane to provide structural support while not interfering with the performance of the antenna. Additionally, a surface-mounted SMA connector with an inner conductor diameter of 1.27 mm is directly soldered to the feed network.

Moreover, the measured results, as shown in Fig. 17, exhibit a similar trend to the simulated ones, with the measured curves fluctuating around the simulated. The measured reflection coefficient within the frequency range of 0.45 GHz to 1.2 GHz is consistently below -10 dB, demonstrating alignment with the simulated one. Although a slight degradation is observed at lower frequencies, the reflection coefficient remains below -8.5 dB. The measured AR closely aligns with the simulated trends within the frequency range of 0.3 GHz to 0.75 GHz, showing a slight degradation in AR at higher frequencies, yet remaining below 3.5 dB. Moreover, the antenna gain demonstrates uniformity across the entire frequency band, with a slight decrease noted at higher frequencies. Differences between measured and simulated values are attributed to errors in assembly and testing, particularly in the connection between the feed network and the antenna, which can affect the phase and impedance. Moreover, it is important to note that the selection of lumped components serves as a source of error. These components not only possess inherent parasitic capacitance, inductance, and resistance but also exhibit frequency-dependent variations in their values. These factors have the potential to alter the phase and impedance, resulting in slight discrepancies between the measured and simulated results.

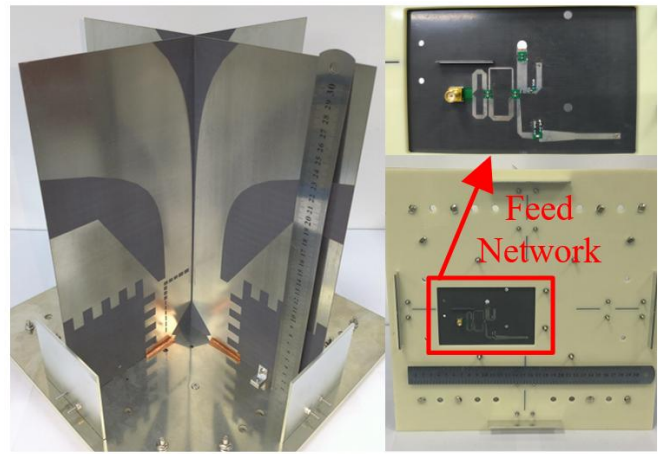

Fig. 16. Photograph of the fabricated antenna.

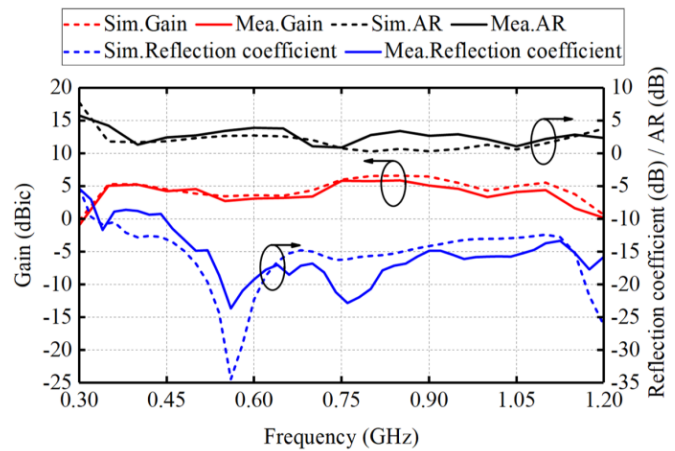

Fig. 17. Measured and simulated reflection coefficients, AR, and Gain of the proposed antenna.

In addition, the measured and simulated normalized patterns of the proposed antenna in *xoz*- and *yoz*-planes at 0.4, 0.7, and 1 GHz are given in Fig. 18(a)-(f). It can be observed that the measured radiation patterns generally agree with the simulated ones.

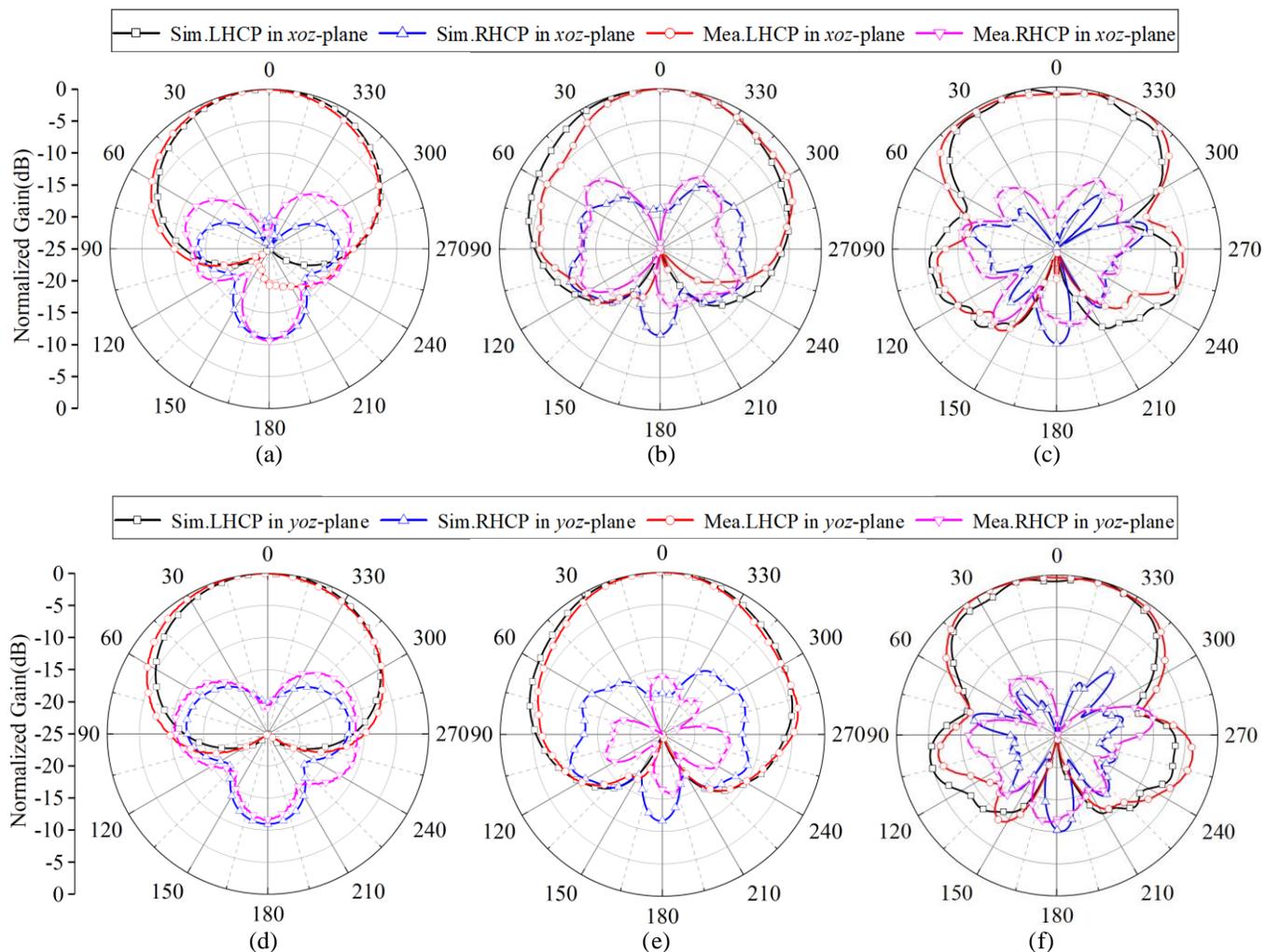

Fig. 18. Measured and simulated radiation patterns in *xoz*-plane at (a) 0.4 Hz, (b) 0.7 GHz, and (c) 1 GHz, in *yoz*-plane at (d) 0.4 GHz, (e) 0.7 GHz, (f) 1 GHz.

TABLE I
COMPARISON BETWEEN REFERENCES AND THIS WORK

| Ref. | Feeding Method | IBW | ARBW | Dimensions ($\lambda_l^2$) | Profile ($\lambda_l$) | Phase Shift Type | The Sizes of The Phase Shifter ($\lambda_l^2$) |
|---|---|---|---|---|---|---|---|
| [15] | Quad-fed | 4.10:1 | 3.61:1 | 1.31 × 2.06 | 4.65 | Directional coupler | 1.31 × 2.06 |
| [16] | Quad-fed | 1.80:1 | 1.52:1 | 0.68 × 0.68 | 0.36 | Loaded-line | 0.59 × 0.59 |
| [17] | Quad-fed | 3.52:1 | 3:1 | 0.30 × 0.30 | 0.61 | Schiffman | 0.29 × 0.24 |
| [18] | Dual-fed | 3.06:1 | 3.9:1 | 0.47 × 0.47 | 0.63 | Loaded-line | 0.42 × 0.25 |
| [19] | Quad-fed | 1.81:1 | 1.68:1 | 0.26 × 0.26 | 0.12 | Loaded-line | 0.26 × 0.26 |
| [20] | Quad-fed | 2.89:1 | 2.35:1 | 0.80 × 0.80 | 0.05 | Schiffman | 0.71 × 0.71 |
| [21] | Dual-fed | 1.68:1 | 1.63:1 | 0.87 × 0.87 | 0.05 | Schiffman | 0.43 × 0.19 |
| [22] | Quad-fed | 2.5:1 | 2.5:1 | 0.59 × 0.59 | 0.06 | Schiffman | 0.52 × 0.40 |
| [23] | Dual-fed | 1.58:1 | 1.54:1 | 0.94 × 0.94 | 0.06 | Loaded-line | 0.54 × 0.27 |
| [24] | Dual-fed | 1.9:1 | 1.89:1 | 0.61 × 0.61 | 0.09 | Loaded-line | 0.38 × 0.32 |
| This work | Dual-fed | 3.75:1 | 3.59:1 | 0.36 × 0.36 | 0.34 | Π-type network And NGD network | 0.071 × 0.047 |

A comprehensive comparison between this work and other proposed wideband CP antenna designs is presented in Table I. The key parameters include the feeding method, 10-dB impedance bandwidth, 3-dB AR bandwidth, antenna dimensions, profile, phase shifter type, and phase shifter sizes. The proposed antenna exhibits ultra-wideband performance, with impedance bandwidth larger than the antennas reported in [16]-[24], and AR bandwidth exceeding that of [16]-[17] and [19]-[24]. Compared to the antenna with a large bandwidth reported in [15], the proposed antenna has significantly reduced dimensions and profile. Although the antenna in [19] has very compact sizes and low profile, its operating bandwidth is significantly narrower than that of the proposed antenna. Additionally, the bandwidth and dimensions of the proposed antenna are superior to those with low profiles [20]-[24]. Notably, the proposed work presents a novel miniaturized phase shifter design employing a Π-type network and an NGD network. Compared to the phase shifter sizes of other antennas in Table I, the one presented in this work has significantly reduced. The proposed phase shifter with sizes of only $0.071 \times 0.047\lambda_l^2$ can be applied to the majority of compact CP antennas.

## VI. CONCLUSIONS

The paper presents a compact ultra-wideband CP antenna based on a novel miniaturized phase shifter. It exhibits a 10-dB impedance bandwidth from 0.32 to 1.2 GHz (3.75:1) and a 3-dB AR bandwidth from 0.32 to 1.15 GHz (3.59:1). Through the utilization of a parasitic radiation structure, sawtooth structures, metal coupled plates, and an Γ-type balun with high-impedance transmission lines, the proposed antenna achieves both miniaturization and broadband impedance matching. The physical dimensions of the antenna are $0.36 \times 0.36 \times 0.34\lambda_l^3$. Additionally, a novel miniaturized phased shifter employing a Π-type network and an NGD network is designed. Compared to conventional ones utilized in CP antennas, the proposed phase shifter achieves a substantial reduction in sizes, with dimensions of only $0.071 \times 0.047\lambda_l^2$. Owing to its minimal dimensions and ultra-wideband characteristics, the proposed phase shifter is applicable to the majority of compact circularly polarized antennas. Finally, a prototype of the design is fabricated and measured to validate its performance, which aligns with simulated results. The excellent performance of the proposed antenna makes it a promising candidate for various UWB applications.

## REFERENCES

[1] D. J. Bisharat, S. Liao and Q. Xue, "Wideband unidirectional circularly polarized antenna with L-shaped radiator structure," *IEEE Antennas and Wireless Propag. Lett.*, vol. 16, pp. 12-15, 2017.
[2] K. M. Mak, H. W. Lai, K. M. Luk and C. H. Chan, "Circularly polarized patch antenna for future 5G mobile phones," *IEEE Access*, vol. 2, pp. 1521-1529, 2014.
[3] J. Wu, Y. Yin, Z. Wang and R. Lian, "Broadband circularly polarized patch antenna with parasitic strips," *IEEE Antennas and Wireless Propag. Lett.*, vol. 14, pp. 559-562, 2015.
[4] Q. W. Lin, H. Wong, X. Y. Zhang and H. W. Lai, "Printed meandering probe-fed circularly polarized patch antenna with wide bandwidth," *IEEE Antennas and Wireless Propag. Lett.*, vol. 13, pp. 654-657, 2014.
[5] M. Li and K. -M. Luk, "A wideband circularly polarized antenna for microwave and millimeter-wave applications," *IEEE Trans. Antennas Propag.*, vol. 62, no. 4, pp. 1872-1879, Apr. 2014.
[6] Y. Li and K. -M. Luk, "A 60-GHz wideband circularly polarized aperture-coupled magneto-electric dipole antenna array," *IEEE Trans. Antennas Propag.*, vol. 64, no. 4, pp. 1325-1333, Apr. 2016.
[7] K. Ding, Y. Li and Y. Wu, "Broadband circularly polarized magnetoelectric dipole antenna by loading parasitic loop," *IEEE Trans. Antennas Propag.*, vol. 70, no. 11, pp. 11085-11090, Nov. 2022.
[8] K. Ding, Y. Li, Y. Li, Y. Wu and J. -F. Li, "Gain-improved wideband circularly polarized magnetoelectric dipole antenna with parasitic helix," *IEEE Trans. Antennas Propag.*, vol. 71, no. 5, pp. 4516-4521, May. 2023.
[9] L. Lu, Y. -C. Jiao, H. Zhang, R. Wang and T. Li, "Wideband circularly polarized antenna with stair-shaped dielectric resonator and open-ended slot ground," *IEEE Antennas and Wireless Propag. Lett.*, vol. 15, pp. 1755-1758, 2016.
[10] N. Yang, K. W. Leung, K. Lu and N. Wu, "Omnidirectional circularly polarized dielectric resonator antenna with logarithmic spiral slots in the ground," *IEEE Trans. Antennas Propag.*, vol. 65, no. 2, pp. 839-844, Feb. 2017.
[11] K. Ding, C. Gao, Y. Wu, D. Qu and B. Zhang, "A broadband circularly polarized printed monopole antenna with parasitic strips," *IEEE Antennas and Wireless Propag. Lett.*, vol. 16, pp. 2509-2512, 2017.
[12] B. Li, S. -W. Liao and Q. Xue, "Omnidirectional circularly polarized antenna combining monopole and loop radiators," *IEEE Antennas and Wireless Propag. Lett.*, vol. 12, pp. 607-610, 2013.
[13] J. Shen, C. Lu, W. Cao, J. Yang and M. Li, "A novel bidirectional antenna with broadband circularly polarized radiation in X-band," *IEEE Antennas and Wireless Propag. Lett.*, vol. 13, pp. 7-10, 2014.
[14] R. Xu, J. -Y. Li, J. -J. Yang, K. Wei and Y. -X. Qi, "A design of U-shaped slot antenna with broadband dual circularly polarized radiation," *IEEE Trans. Antennas Propag.*, vol. 65, no. 6, pp. 3217-3220, Jun. 2017.
[15] J. -C. Liang, C. -N. Chiu, T. -C. Lin and C. -H. Lee, "An ultrawideband circularly-polarized Vivaldi antenna with high gain," *IEEE Access*, vol. 10, pp. 100446-100455, 2022.
[16] Y. -J. Hu, Z. -M. Qiu, B. Yang, S. -J. Shi and J. -J. Yang, "Design of novel wideband circularly polarized antenna based on Vivaldi antenna structure," *IEEE Antennas and Wireless Propag. Lett.*, vol. 14, pp. 1662-1665, 2015.
[17] X. -Y. Sun et al., "Ultrawideband circularly polarized halved-type Vivaldi antenna with symmetrical radiation pattern," *IEEE Antennas and Wireless Propag. Lett.*, vol. 23, no. 2, pp. 633-637, Feb. 2024.
[18] X. Ren, S. Liao and Q. Xue, "Design of wideband circularly polarized Vivaldi antenna with stable radiation pattern," *IEEE Access*, vol. 6, pp. 637-644, 2018.
[19] X. Liu, Y. Zhu and W. Xie, "A miniaturized wideband directional circularly polarized antenna based on bent Vivaldi antenna structure," *IEEE Antennas and Wireless Propag. Lett.*, vol. 22, no. 2, pp. 298-302, Feb. 2023.
[20] L. Gu, W. Yang, W. Feng, Q. Xue, Q. Meng and W. Che, "Low-profile ultrawideband circularly polarized metasurface antenna array," *IEEE Antennas and Wireless Propag. Lett.*, vol. 19, no. 10, pp. 1714-1718, Oct. 2020.
[21] L. -L. Qiu, L. Zhu and Y. Xu, "Wideband low-profile circularly polarized patch antenna using 90° modified Schiffman phase shifter and meandering microstrip feed," *IEEE Trans. Antennas Propag.*, vol. 68, no. 7, pp. 5680-5685, Jul. 2020.
[22] C. Sun, "A design of compact ultrawideband circularly polarized microstrip patch antenna," *IEEE Trans. Antennas Propag.*, vol. 67, no. 9, pp. 6170-6175, Sep. 2019.
[23] Z. -H. Wu, Y. Lou and E. K. -N. Yung, "A circular patch fed by a switch line balun with printed L-probes for broadband cp performance," *IEEE Antennas and Wireless Propag. Lett.*, vol. 6, pp. 608-611, 2007.
[24] J. Zhuang, Y. Zhang, W. Hong and Z. Hao, "A broadband circularly polarized patch antenna with improved axial ratio," *IEEE Antennas and Wireless Propag. Lett.*, vol. 14, pp. 1180-1183, 2015.
[25] H. Mirzaei and G. V. Eleftheriades, "Realizing non-foster reactive elements using negative-group-delay networks," *IEEE Trans. Microw. Theory Techn.*, vol. 61, no. 12, pp. 4322-4332, Dec. 2013.